\documentclass{aastex63}

\received{?}
\revised{\today}
\accepted{August 10, 2020}
\submitjournal{ApJ}
\shorttitle{Pulsations of KIC~10685175}
\shortauthors{Shi et al.}
\graphicspath{{./}{figures/}}
\usepackage{natbib}
\usepackage{subfigure}
\usepackage{color}

\begin{document}
\title{Pulsations of the roAp star KIC~10685175 revisited by TESS}

\author{Fangfei Shi}
\affiliation{Department of Astronomy, School of Physics, Peking University, Beijing 100871, P. R. China}
\affiliation{Kavli Institute for Astronomy and Astrophysics, Peking University, Beijing 100871, P. R. China}

\author{Donald Kurtz}
\affiliation{Centre for Space Research, Physics Department, North West University, Mahikeng 2745, South Africa}
\affiliation{Jeremiah Horrocks Institute, University of Central Lancashire, Preston PR1~2HE, UK}

\author{Hideyuki Saio}
\affiliation{Astronomical Institute, Graduate School of Science, Tohoku University, Sendai 980-8578, Japan}

\author{Jianning Fu}
\affiliation{Department of Astronomy, Beijing Normal University, Beijing, 100871, P. R. China}

\author{Huawei Zhang}
\affiliation{Department of Astronomy, School of Physics, Peking University, Beijing 100871, P. R. China}
\affiliation{Kavli Institute for Astronomy and Astrophysics, Peking University, Beijing 100871, P. R. China}

\begin{abstract}
KIC~10685175 (TIC~264509538) was discovered to be a rapidly oscillating Ap star from {\it Kepler} long cadence data using super-Nyquist frequency analysis. It was re-observed by TESS with 2-min cadence data in Sectors 14 and 15. We analyzed the TESS light curves, finding that the previously determined frequency is a Nyquist alias. The revised pulsation frequency is $191.5151 \pm 0.0005$\,d$^{-1}$ ($P = 7.52$\,min) and the rotation frequency is $0.32229 \pm 0.00005$\,d$^{-1}$ ($P_{\rm rot} = 3.1028$\,d). The star is an oblique pulsator with pulsation amplitude modulated by the rotation, reaching pulsation amplitude maximum at the time of the rotational light minimum. The oblique pulsation generates a frequency quintuplet split by exactly the rotation frequency. The phases of sidelobes, the pulsation phase modulation, and a spherical harmonic decomposition all show this star to be pulsating in a distorted quadrupole mode. Following the oblique pulsator model, we calculated the rotation inclination $i$ and magnetic oblique $\beta$ of this star, which provide detailed information of pulsation geometry. The $i$ and $\beta$ derived by the best fit of pulsation amplitude and phase modulation through a theoretical model differ from those calculated for a pure quadrupole, indicating the existence of strong magnetic distortion. The model also predicts the polar magnetic field strength is as high as about 6\,kG which is predicted to be observed in a high resolution spectrum of this star. 
\end{abstract}

\keywords{Stellar astronomy: Rapid stellar oscillations - Stellar astronomy: Oblique rotators}

\section{Introduction} \label{sec:intro}

The chemically peculiar A (Ap)  stars have non-uniform distributions of chemical abundances on their surfaces and strong magnetic fields. Some cool Ap stars exhibit high-overtone, low-degree pressure pulsation modes with periods between 4.7 and 24\,min and amplitudes up to 0.018 mag in Johnson $B$ \citep{2019MNRAS.487.3523C,2009CoAst.159...61K,2015MNRAS.452.3334S}. They are called rapidly oscillating Ap (roAp) stars. The strong magnetic fields of roAp stars can suppress convection leading to stratification, so that some rare earth elements,  such as Eu, Pr, and Nd, have enhanced spectral lines through radiative levitation, hence show over-abundances. These elemental overabundances occur in spots, making Ap stars, in general, and roAp stars, in particular, obliquely rotating variable stars of a class known as $\alpha^2$~CVn stars \citep{1969ApJS...18..347P}.

\citet{1950MNRAS.110..395S} developed the oblique rotator model of the Ap stars, which accounts well for the form of the magnetic, spectrum, and light variations. Following this model, \citet{1982MNRAS.200..807K} introduced the oblique pulsator model, which was generalized with the effects of both the magnetic field and rotation taken into account \citep{1982MNRAS.200..807K,1985ApJ...296L..27D,1993PASJ...45..617S,1994PASJ...46..301T,1995PASJ...47..219T,2004MNRAS.350..485S,2002A&A...391..235B,2011A&A...536A..73B}. According to this model, the pulsation axis is misaligned with the rotation axis, and generally close to the magnetic axis. When the star rotates, different pulsation aspect is seen along the line of sight, leading to observed amplitude and phase modulation. This modulation can provide information on the geometry of observed pulsations, hence mode identification, which is necessary for asteroseismic inference with forward modelling.

Since the first roAp stars were discovered by \citet{1982MNRAS.200..807K}, 78 rapidly oscillating Ap (roAp) stars have been found \citep{2015MNRAS.452.3334S, 2019MNRAS.488...18H,2019MNRAS.487.3523C,2019MNRAS.487.2117B}  Asteroseismology is a useful method to diagnose stellar structure and interior physics from the evidence of surface pulsations. Progress of this research for roAp stars has been hindered by the relatively small number of known stars, and because their rapid pulsation requires dedicated observations at a short enough cadence \citep{2019MNRAS.488...18H,2019MNRAS.487.3523C,2019MNRAS.487.2117B}. 

Fortunately, the space telescopes {\it Kepler} and TESS provide us with continuous light curves, and the spectroscopic telescope LAMOST provides us with spectra to identify chemically peculiar stars. Although the sampling frequency of the {\it Kepler} long cadence data is lower than the typical pulsation frequencies of roAp stars, \citet{2013MNRAS.430.2986M} showed that there is no limitation of the Nyquist frequency in the {\it Kepler} long cadence data because of the periodically modulated time-sampling. That is, when the sampling interval is longer than half the pulsation period, in the amplitude spectrum the Nyquist aliases are split into multiplets while the real frequency is a single peak, so that it is distinguishable from the aliases. \citet{2019MNRAS.488...18H} found six new roAp stars through this method. 

Our target, KIC~10685175 (TIC~264509538), is among the six stars found in {\it Kepler} long cadence data \citep{2019MNRAS.488...18H}. In their work, the pulsation frequency was found to be 240.45189\,d$^{-1}$ (2783.0080\,$\mu$Hz), which we show is a Nyquist alias, and the rotation frequency was found to be $0.32237 \pm 0.00001$\,d$^{-1}$ (3.7311\,$\mu$Hz). The effective temperature $T_{\rm eff} = 8000\pm300$\,K was obtained from LAMOST, the magnitude $g = 12.011$ from SDSS, and the luminosity $\log(L/ {\rm L}_\odot) = 0.896\pm0.022$ from Gaia DR2. The mass of this star was estimated to be $M = 1.65\pm0.25$\,M$_{\odot}$ through an interpolation over stellar tracks based on the models of \citet{2013MNRAS.436.1639C}. 

\section{TESS observations} \label{sec:rotation}

\begin{figure}[ht!]
\centering
        \includegraphics[scale=0.3]{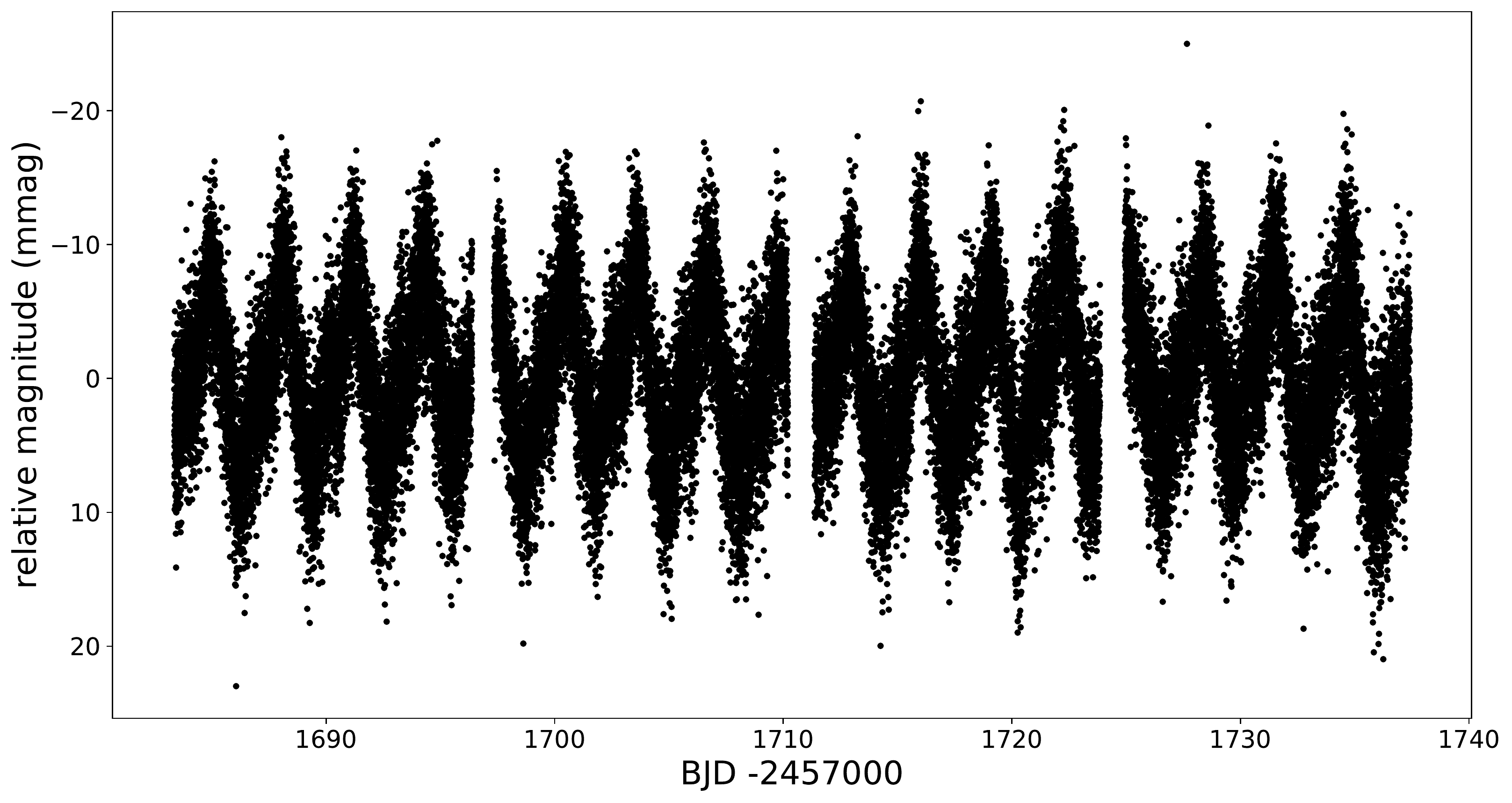}
        \includegraphics[scale=0.3]{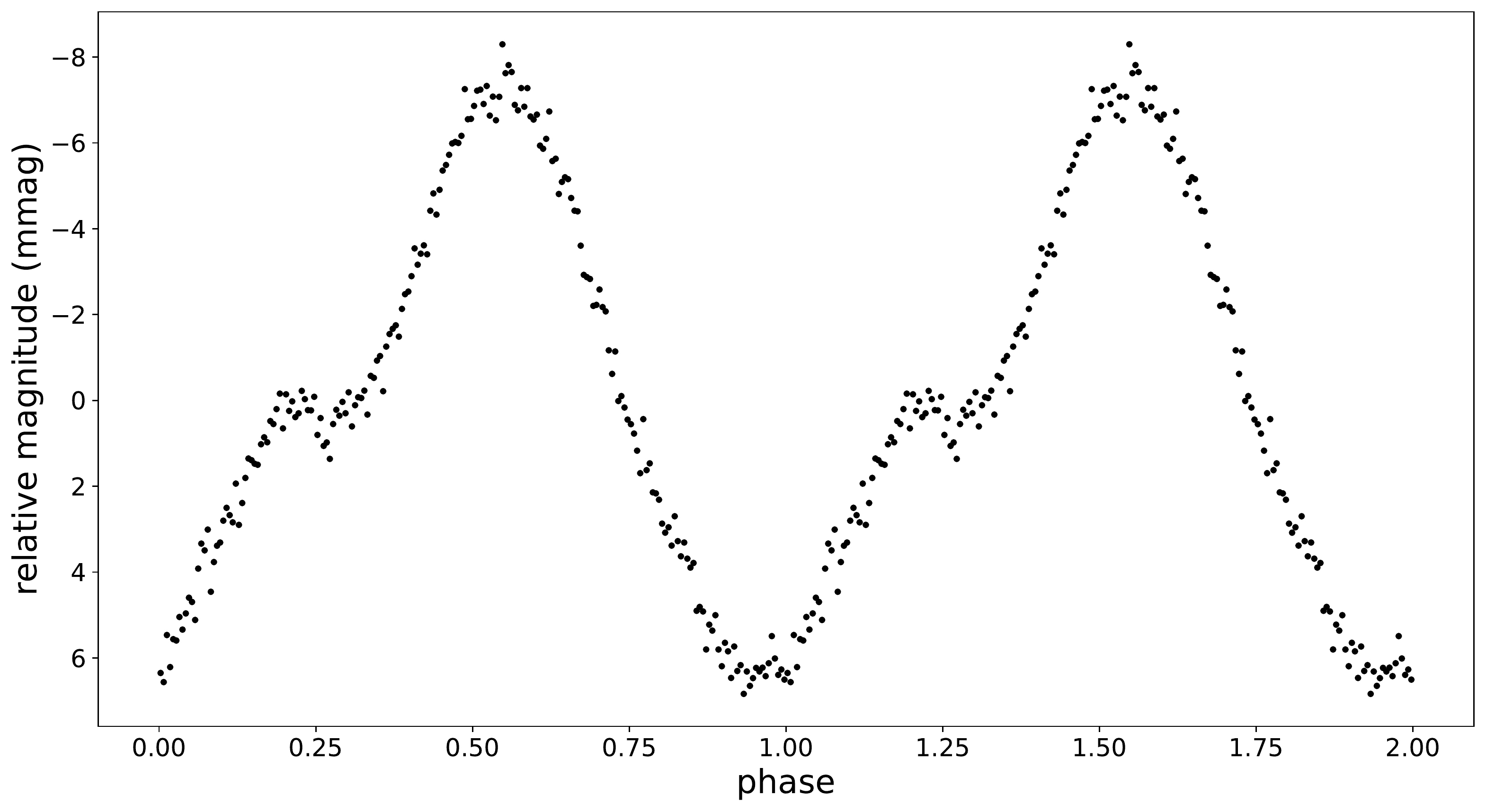}
\caption{\label{1} The light curve (top) and phase folded light curve (bottom) of KIC~10685175, folded on the rotation period of 3.10198\,d; two rotation cycles are shown for clarity. The data are from TESS sectors 14 and 15. The time zero-point is BJD\,2458711.21391.}
\label{fig:lc}
\end{figure}

The Transiting Exoplanet Survey Satellite \citep[TESS, ][]{2015JATIS...1a4003R} is aimed at detecting planets through the transit method, and asteroseismology also benefits greatly from the high precision photometric data. KIC~10685175 was observed by TESS with 2-min cadence in Sectors 14 and 15. The light curve contains 38094 data points from BJD~2458476.95 to 2458531.01, for a total time span of 54.06\,d. After removing bad data points, the standard PDC$\_$SAP flux produced by the Science Processing Operations Center was used. The original light curve shows some time gaps and has different mean fluxes between the two sectors. The fluxes were corrected by dividing by the median flux separately for each sector. Relative magnitudes were then calculated from the processed fluxes, giving the light curve shown in the top panel of Figure\,\ref{fig:lc}.

\section{Frequency analysis} \label{sec:frequency}
\subsection{Rotation frequency analysis}

\begin{figure}[ht!]
\centering
        \includegraphics[scale=0.3]{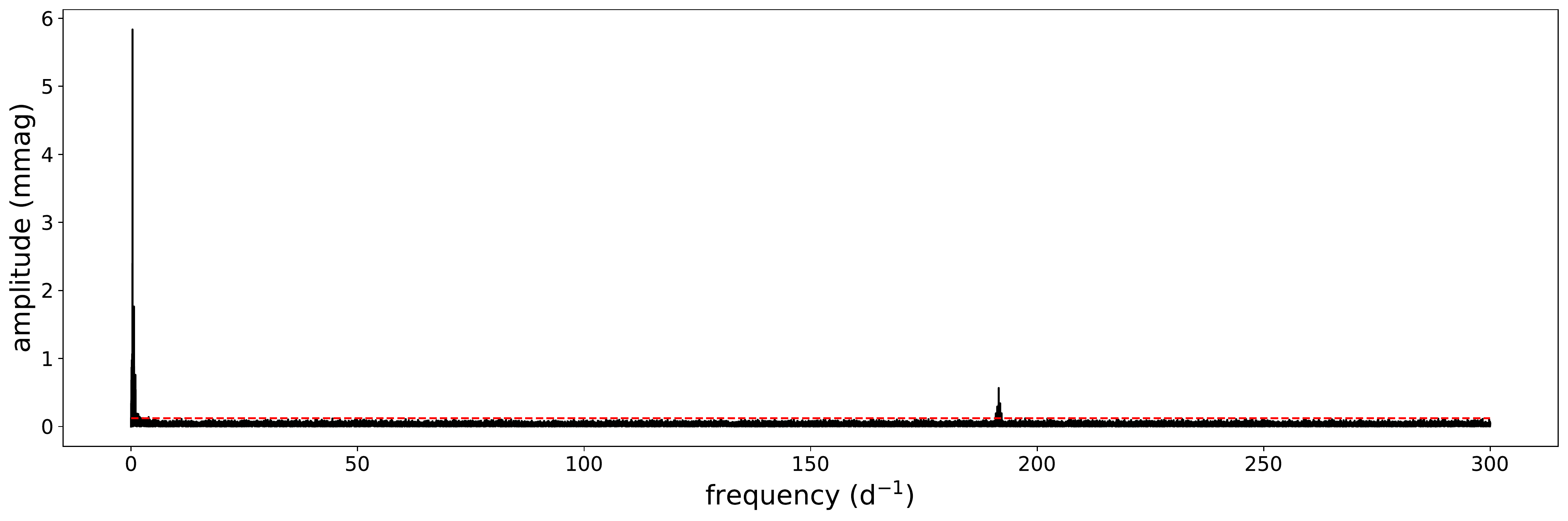}
        \includegraphics[scale=0.3]{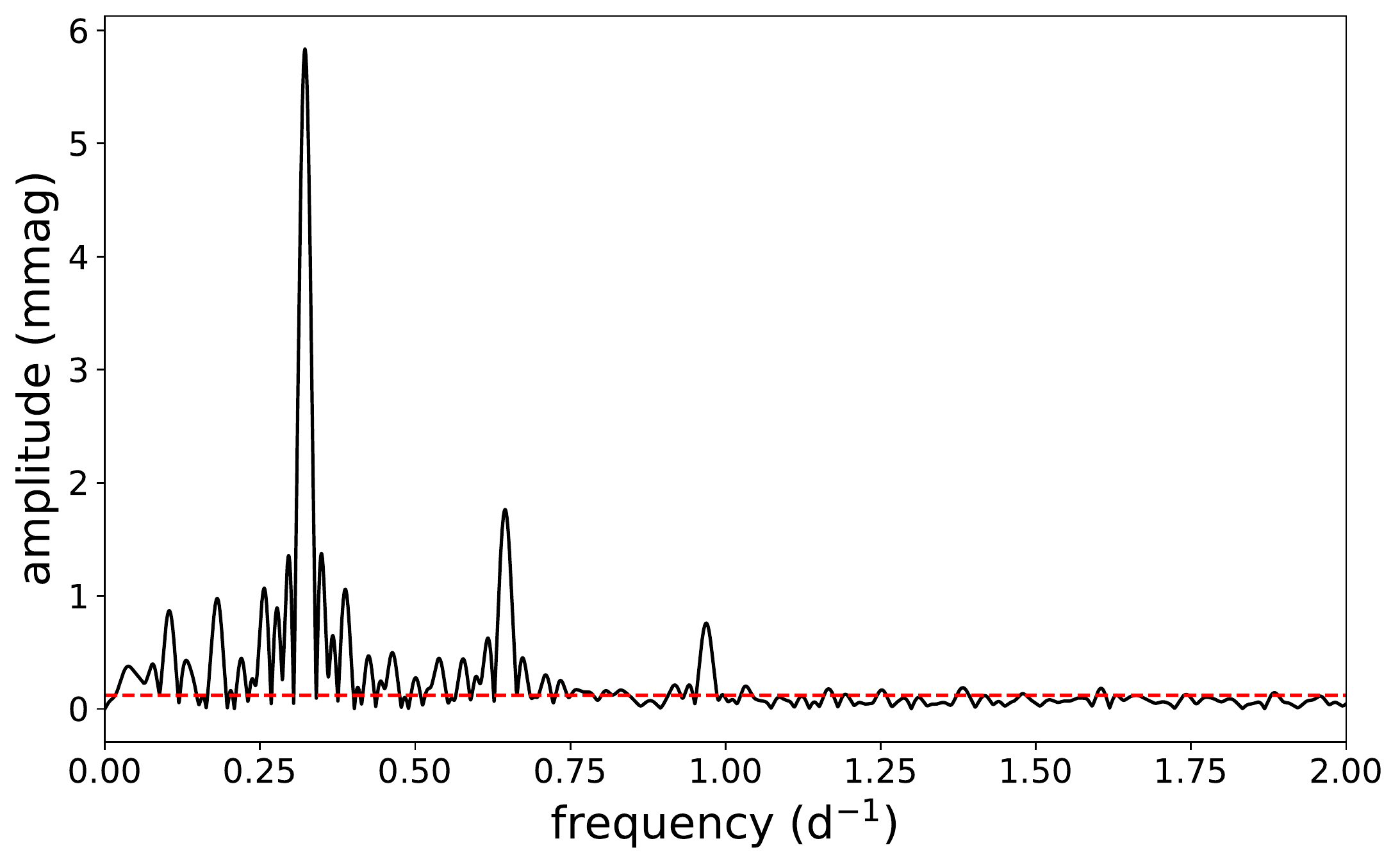}
        \includegraphics[scale=0.3]{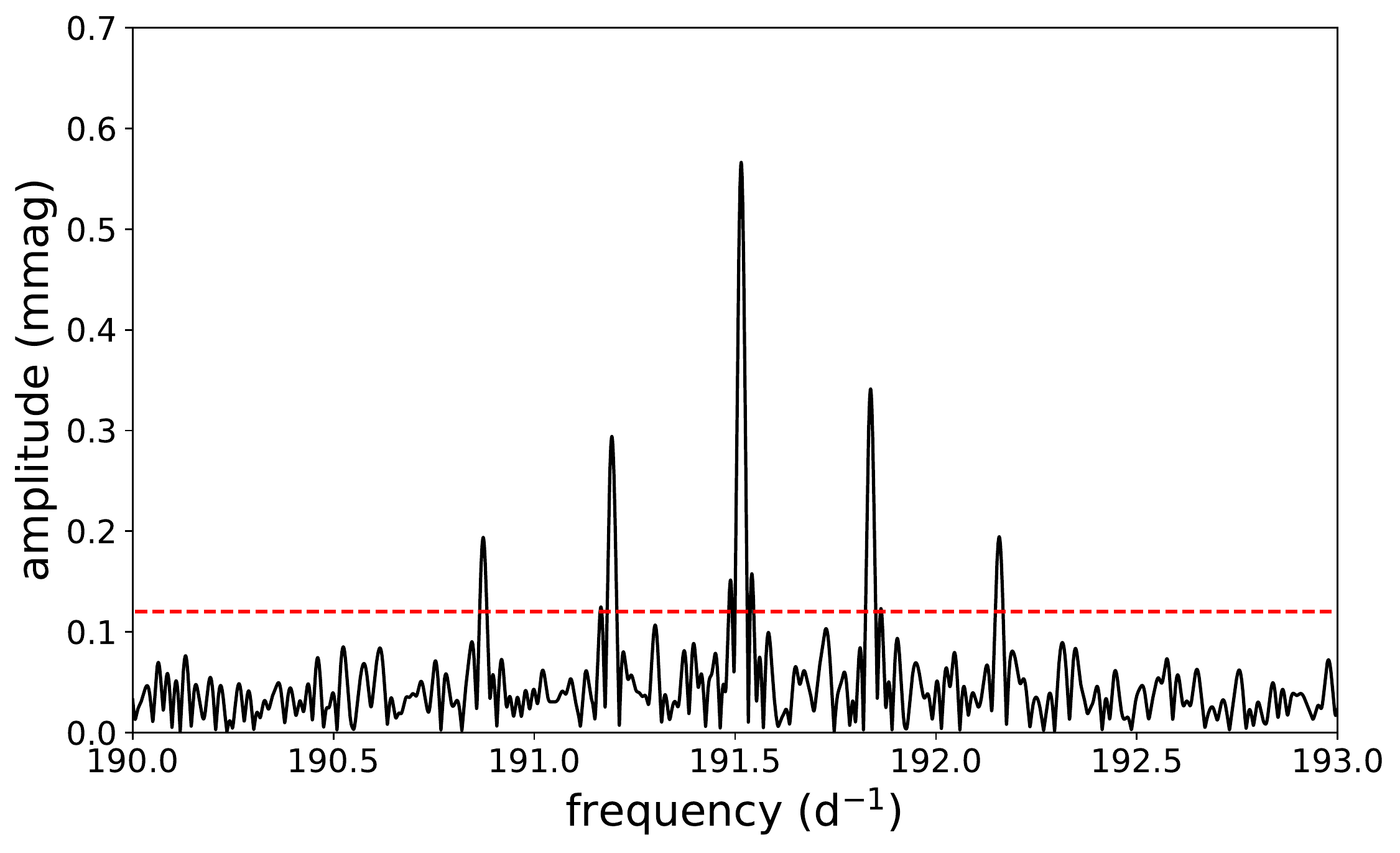}
\caption{\label{2} The top panel shows the amplitude spectrum of KIC~10685175; the left and right bottom panels show higher frequency resolution views of the two sections where the rotation (left) and pulsation (right) signals are seen; Note the different ordinate scales on these two panels.  The red line is 4.6 times the background signal. According to \citet{2016A&A...585A..22Z}, for space data, it is better to take higher threshold above which a signal can safely be considered as real, so here 4.6 times the background signal is used.}
\label{fig:ft}
\end{figure}

There is an obvious rotation signal in the light curve caused by surface spots. These are inclined to the rotation axis, such that when the spots cross the line of sight, an extremum is seen in the flux. In Ap stars this can be a maximum or a minimum, depending on how the spots redistribute the flux as a function of wavelength (in practice, the observational passband). For many Ap stars a maximum in brightness is seen in the red when there is a minimum in the blue; see, e.g., \citet{1996MNRAS.280....1K} for the roAp star HD~6532 and \citet{1990MNRAS.243..289K} for the roAp star HD~60435. This is a consequence of line blocking increasing the photospheric temperature gradient, hence redistributing flux from blue to red because of the different depths probed. The flux variation from the spots is stable so that it is possible to determine the rotation period precisely. We calculated the amplitude spectrum of the data in magnitudes, shown in the upper left panel of Figure\,\ref{fig:ft}, from which we derived the rotation frequency to be $\nu_{\rm rot} =0.32229 \pm 0.00005$\,d$^{-1}$ ($P_{\rm rot} = 3.1028 \pm 0.0005$\,d). This rotation period is consistent with previous work ($P_{\rm rot} = 3.10198 \pm 0.00001$\,d) determined by \citet{2019MNRAS.488...18H} from 4\,yr of {\it Kepler} data. Since the data length of TESS data is much shorter than that of {\it Kepler}, and the TESS data is lower precision, our rotation frequency uncertainty is larger. The phase folded light curve after removing pulsation signals and binning to 0.005 phase is shown in the bottom panel of Figure\,\ref{fig:lc}. The double wave suggests that both pulsation poles of the star are seen over the rotation cycle, under the reasonable assumption for an Ap star that there are spots at, or near to, both poles.

\subsection{Pulsation frequency analysis}

Along with the low frequency peaks from rotation and instrumental effects in the periodogram, there is a group of obvious high frequencies around 191\,d$^{-1}$ which come from pulsation in the roAp star. The higher amplitude low frequency peaks contribute to the total variance in the data, and their spectral window functions extend out to the pulsation frequencies. To remove these effects of the rotation frequencies and some instrumental drifts, we ran a high-pass filter by first pre-whitening the rotation frequency and its second and third harmonics, then dividing the light curve into 124 segments for every 300 data points. Polynomials were then fitted and removed from these segments to remove both the rotation and instrumental effects clearly.  

In the periodogram of the light curve after the high-pass filter was applied, a quintuplet of frequencies is seen centred on $\nu_{1}$=191.5153\,d$^{-1}$ (Figure\,\ref{fig:ft}, bottom right panel). The principal frequency and its rotational sidelobes at $\nu_{1} \pm \nu_{rot}$ and $\nu_{1} \pm 2\nu_{rot}$ are clear. A non-linear least squares fit of the rotation frequency, two of its harmonics, and the pulsation quintuplet  is shown in Table~1. The results show clearly that the quintuplet frequencies are split by exactly the rotation frequency within the uncertainties. Thus, the frequency quintuplet sidelobes were next fixed to be equally spaced by the rotation frequency according to the oblique pulsator model, and zero-point in time was chosen such that the phases of the first pair of sidelobes are the same, then a linear least squares fit was applied to the data with the results shown in Table~2. Within the uncertainties, the phases of all five members of the quintuplet are equal, although the amplitudes are not completely symmetric, as a result of the magnetic perturbation. This thus suggests a distorted quadrupole mode, as a pure quadrupole mode would have symmetric amplitudes and equal phases. 
There are now 7 roAp stars \citep{2019MNRAS.489.4063H,2018MNRAS.476..601H,2018MNRAS.480.2405H,2018MNRAS.473...91H,2014MNRAS.443.2049H,1996MNRAS.281..883K,2017EPJWC.16003004H} for which quadrupole modes have been found, and all of them have distorted modes.

According to the oblique pulsator model, the amplitudes of the rotation sidelobes of the pulsation frequency depend on the star's rotation inclination ($i$) and  magnetic obliquity ($\beta$) angles. For a quadrupole mode we can derive a constraint on $i$ and $\beta$ from \citep{1990ARA&A..28..607K}:
\begin{equation}
\tan i \tan\beta=4\frac{A^{(2)}_{+2}+A^{(2)}_{-2}}{A^{(2)}_{+1}+A^{(2)}_{-1}},
\end{equation}
where $A^{(2)}_{-2}$, $A^{(2)}_{-1}$, $A^{(2)}_{+1}$, $A^{(2)}_{+2}$ are the amplitudes of pulsation frequency sidelobes given in Table~2, which thus give $\tan i \tan\beta=2.40 \pm 0.30$. The value is consistent with the value of $\tan i \tan\beta=1.7 \pm 1.6$ obtained by \citet{2019MNRAS.488...18H} from {\it Kepler} long cadence data, but with much smaller uncertainties because of the shorter cadence of the TESS observations.

\begin{deluxetable*}{cccc}
\tablenum{1}
\tablecaption{A non-linear least squares fit to the rotation frequency and its harmonics, and the pulsation frequency and its rotational sidelobes for KIC~10685175. The zero-point for the fit is BJD\,2458711.21931. }
\tablewidth{0pt}
\tablehead{
\colhead{  } & \colhead{Frequency } & \colhead{Amplitude } & \colhead{Phase}\\
\colhead{  } & \colhead{(d$^{-1}$)} & \colhead{(mmag)} & \colhead{(rad)} \\
\colhead{  } & \colhead{} & \colhead{($\pm 0.028$)} & \colhead{}
}
\startdata
$\nu_{rot}$ & $0.32229 \pm 0.00005$ & $5.806$ & $6.237 \pm 0.005$\\
$2\nu_{rot}$ & $0.6451 \pm 0.0002$ & $1.791 $ & $1.75 \pm 0.02$ \\
$3\nu_{rot}$ & $0.9668 \pm 0.0004$ & $0.710 $ & $5.97 \pm 0.04$ \\
$\nu_{1}-2\nu_{rot}$ & $190.8735 \pm 0.0014$ & $0.198 $ & $0.52 \pm 0.14$ \\
$\nu_{1}-\nu_{rot}$ & $191.1935 \pm 0.0010$ & $0.291 $ & $0.44 \pm 0.10$\\
$\nu_{1}$ & $191.5151 \pm 0.0005$ & $0.565 $ & $0.73 \pm 0.05$\\
$\nu_{1}+\nu_{rot}$ & $191.8369 \pm 0.0008$ & $0.341 $ & $0.45 \pm 0.08$\\
$\nu_{1}+2\nu_{rot}$ & $192.1569 \pm 0.0014$ & $0.196 $ & $0.78 \pm 0.14$\\
\enddata
\end{deluxetable*}

\begin{deluxetable*}{cccc}
\tablenum{2}
\tablecaption{Linear least squares fit to the pulsation frequency and rotational sidelobes forced to have exact splitting of the rotation frequency for KIC~10685175. The zero-point for the fit is BJD\,2458711.21931.}
\tablewidth{0pt}
\tablehead{
\colhead{ Flag } & \colhead{Frequency} & \colhead{Amplitude} & \colhead{Phase}\\
\colhead{  } & \colhead{(d$^{-1}$)} & \colhead{(mmag)} & \colhead{(rad)}\\
\colhead{  } & \colhead{} & \colhead{($\pm 0.028$)} & \colhead{}
}
\startdata
$\nu_{1}-2\nu_{rot}$ & 190.8705 & $0.189 $ & $0.52 \pm 0.15$\\
$\nu_{1}-\nu_{rot}$ & 191.1928 & $0.290 $ & $0.45 \pm 0.08$\\
$\nu_{1}$ & 191.5151 & $0.563 $ & $0.73 \pm 0.05$\\
$\nu_{1}+\nu_{rot}$ & 191.8374 & $0.339 $ & $0.45 \pm 0.10$\\
$\nu_{1}+2\nu_{rot}$ & 192.1597 & $0.189 $ & $0.79 \pm 0.15$\\
\enddata
\end{deluxetable*}

\subsection{Pulsation amplitude and phase modulation}

\begin{figure}[ht!]
\centering
        \includegraphics[scale=0.3]{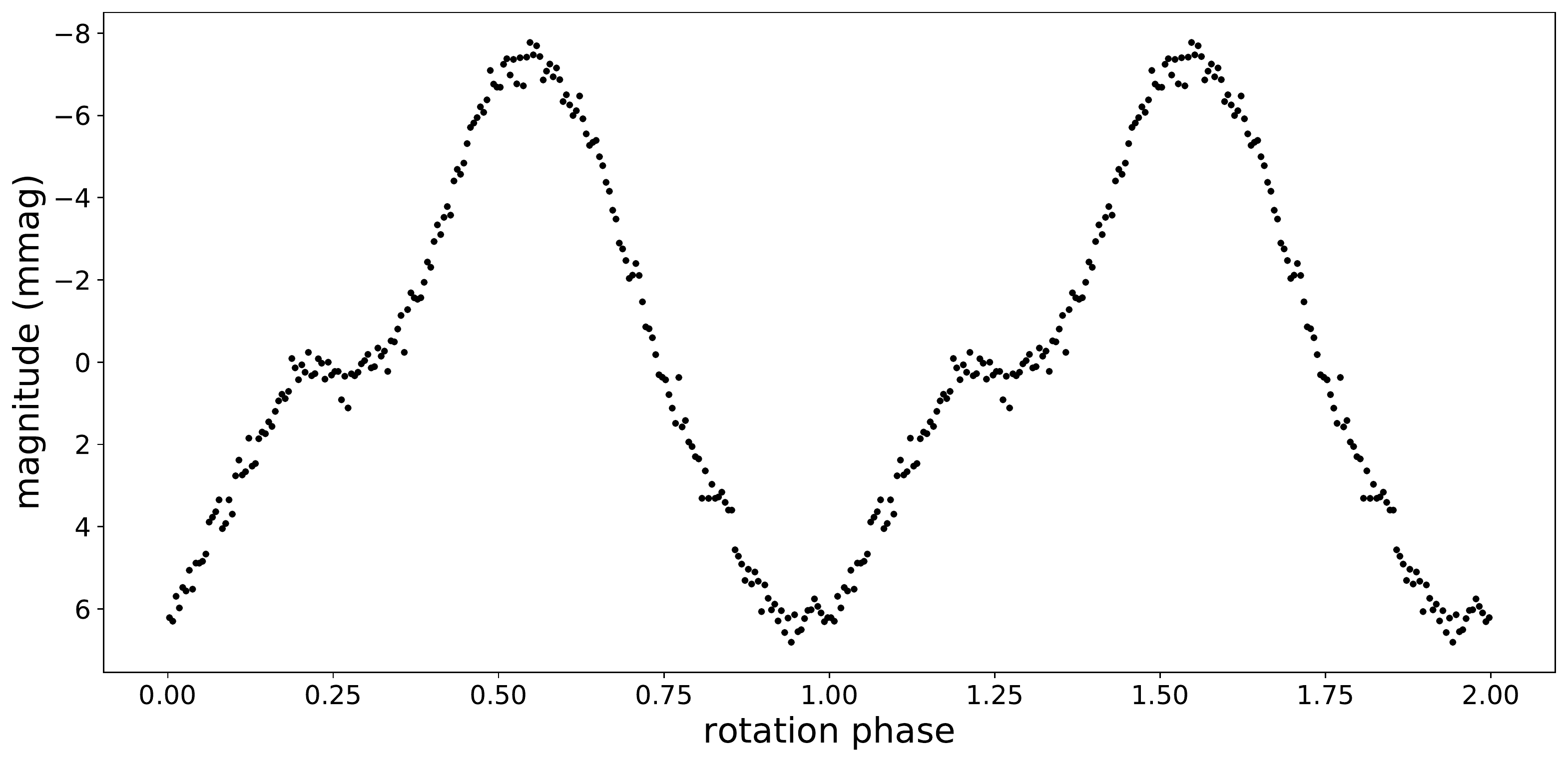}
        \includegraphics[scale=0.3]{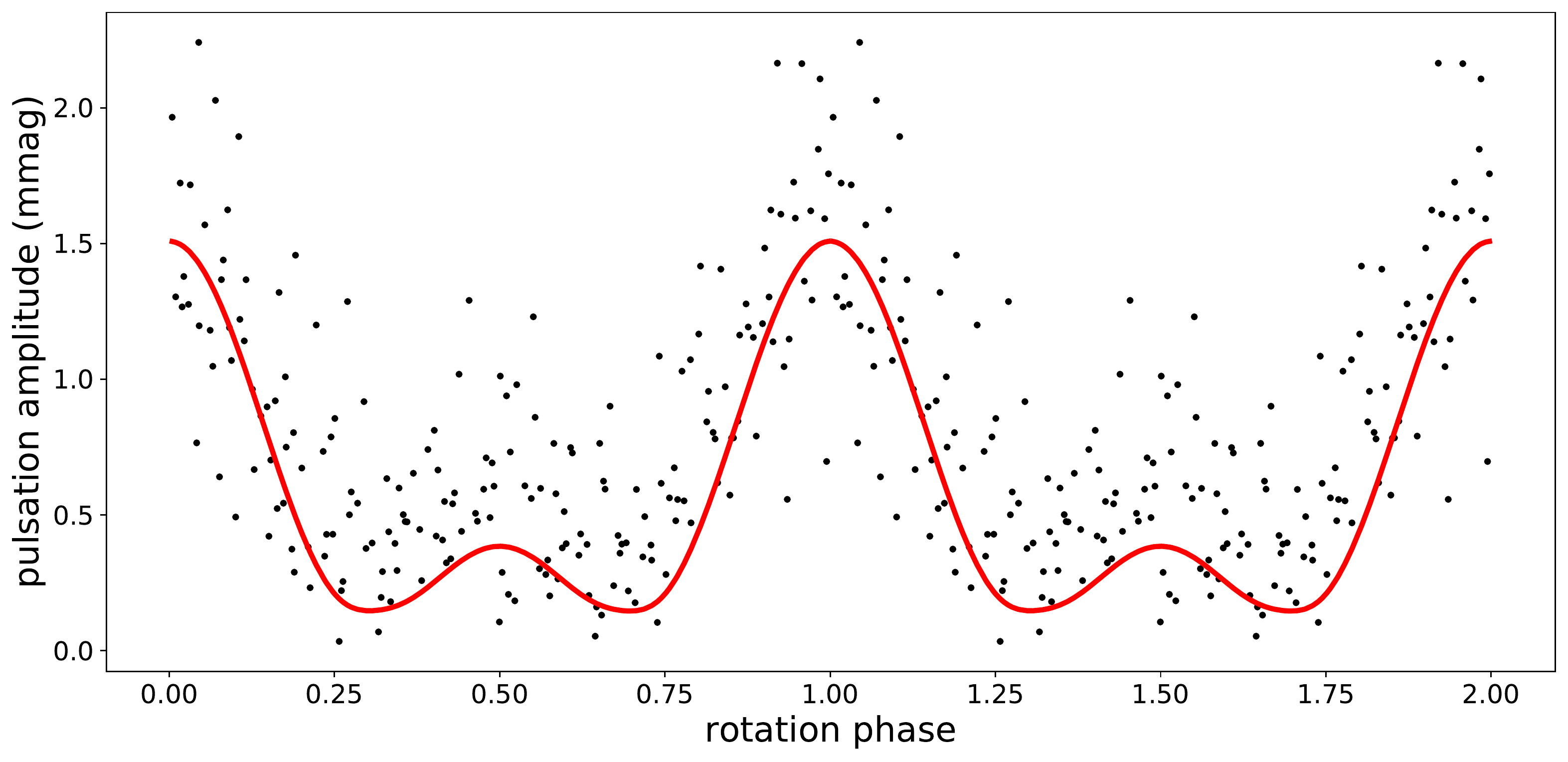}
        \includegraphics[scale=0.3]{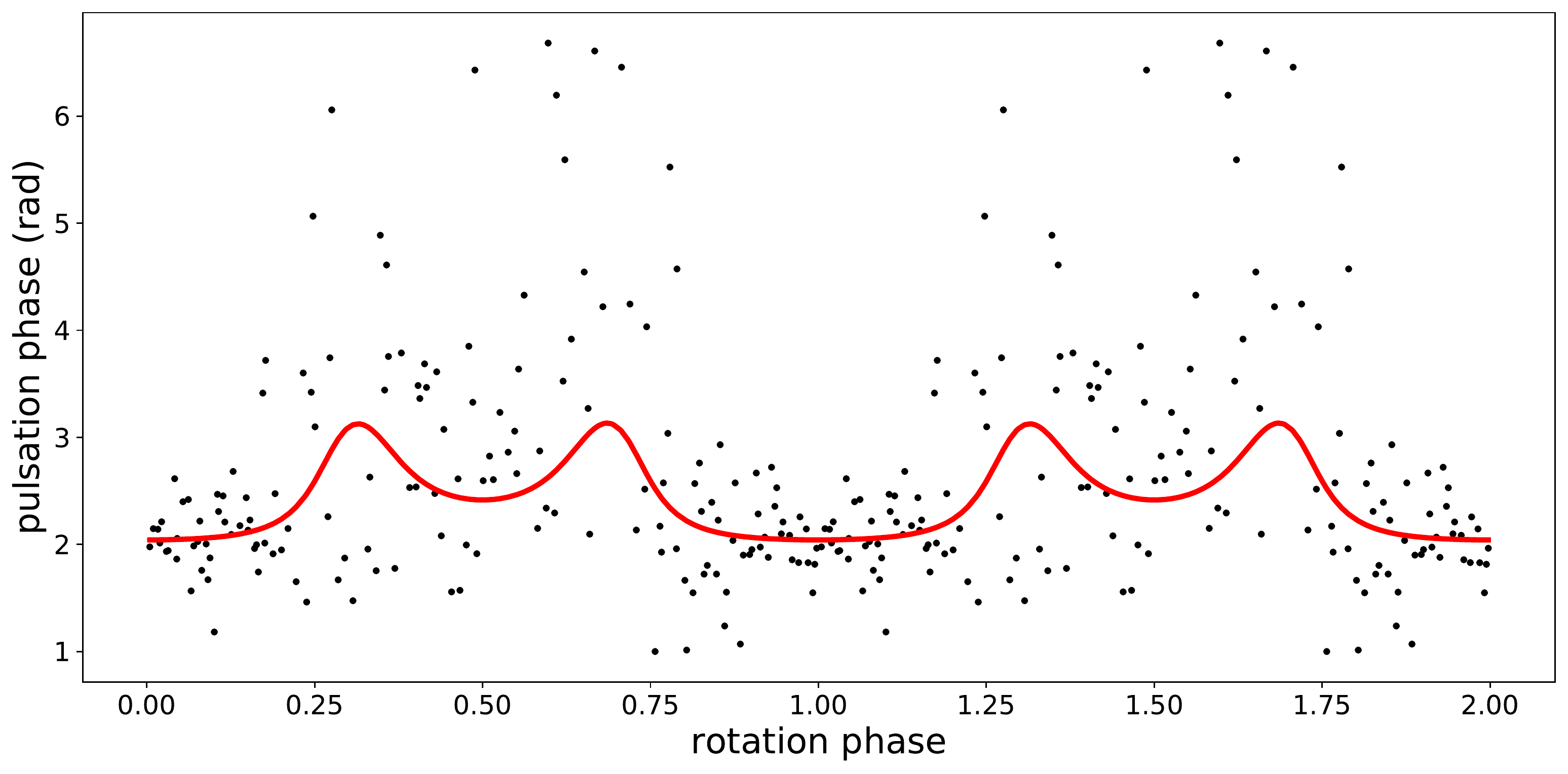}
\caption{\label{3} The phase folded rotation light curve is shown (top) for comparison with the pulsation amplitude (middle) and phase (bottom) variations as a function of rotation phase. The errors of the phases are large around the minimum of the pulsation amplitude, since the error on phase is proportional to the signal-to-noise ratio in the amplitude. A few phase points with {$1\sigma$} errors greater than 1\,rad deviate strongly from others which will smooth the bump, so these outlying points are not plotted here.} The red points are theoretical amplitude modulation modelled following \citet{1992MNRAS.259..701K}. Two rotation cycles are shown. The time zero-point is BJD\,2458711.21391.
\label{fig:modulation}
\end{figure}

To study the rotation modulation of the pulsation amplitudes and phases, the light curve was divided into 196 segments each containing 50 pulsation cycles, thus each segment had a time span of 0.26\,d, or 0.08 of a rotation cycle. Linear least-squares fitting was applied to these segments at fixed frequency, $\nu_{1} = 191.5151$\,d$^{-1}$. The pulsation amplitude and phase variations as a function of rotational phase, along with the rotation light variations for comparison, are shown in Figure\,\ref{fig:modulation}. At rotation phase 0, the pulsation amplitude peaks when the rotation light is minimum. For roAp stars, spots are assumed to be situated close to the pulsation axis, so the pulsation maximum coincides with rotation extremum. Whether the extremum is maximum or minimum depends on how the spots redistribute the surface energy. Usually, for other roAp stars, pulsation maximum coincides with rotational minimum light in a blue filter and maximum light in a red filter, as mentioned in Section 3.1 above. The TESS filter is redder than $V$, so this unusual result suggests that the rotational light curve of KIC~10685175 is probably in phase in both blue and red passbands. This should be tested by observing this star in different passbands to check if there any phase differences, just as the work in \citep{Handler_2006}.

For a pure quadrupole pulsator, the pulsation amplitude peaks at the two pulsation poles and equator, and the pulsation amplitude variation over the star follows the second Legendre polynomial, $\frac{1}{2}(3\cos^2\theta -1)$, where $\theta$ is co-latitude, the angle to the poles. The pulsation maximum at the poles is twice that at the equator, but with inverse phase. In Figure\,\ref{fig:modulation} it can be seen that the pulsation amplitude shows a double wave over the rotation cycle, with two maxima of unequal heights. This suggests a rotational inclination $i$ and a magnetic obliquity $\beta$ such that the higher maximum is dominated by the pulsation polar cap, and the other is dominated by the pulsation equator. The amplitude should go to zero for a pure quadrupole mode when the line of sight crosses a surface pulsation node. In the case of KIC~10685175 it can be seen that the pulsation amplitude drops, but not quite to zero. This is a consequence of the mode being distorted, rather than a pure quadrupole mode. The shape of the amplitude modulation curve in the middle panel of Figure~3 suggests that pulsation maximum occurs when the pulsation pole is closest to the line of sight. Then with rotation, the line of sight crosses one surface node so that the pulsation equatorial belt dominates and the amplitude rises to the secondary maximum. However, the pulsation phase as a function of rotation 
does not show a $\pi$-rad phase reversal expected at the times of amplitude minima, although the pulsation phase is perturbed at those times. This then argues for a distorted quadrupole mode, and this is similar to what is observed in other the roAp stars with well-studied quadrupole modes \citep{2019MNRAS.489.4063H,2018MNRAS.476..601H,2018MNRAS.480.2405H,2018MNRAS.473...91H,2014MNRAS.443.2049H,1996MNRAS.281..883K,2017EPJWC.16003004H}.  

We now look at the geometry of the mode quantitatively. From the oblique pulsator model \citep{1990MNRAS.247..558K} the rotation inclination $i$ and magnetic oblique $\beta$ of a quadrupole pulsator can be derived from the amplitude ratios of the sidelobes through:
\begin{equation}
\frac{A^{(2)}_{+2}+A^{(2)}_{-2}}{A^{(2)}_{0}}=\frac{3\sin^{2}\beta \sin^{2}i}{(3\cos^{2}\beta -1)(3\cos^{2}i -1)} 
\end{equation}
\begin{equation}
\frac{A^{(2)}_{+1}+A^{(2)}_{-1}}{A^{(2)}_{0}}=\frac{12\sin\beta \sin i \cos\beta \cos i}{(3\cos^{2}\beta -1)(3\cos^{2}i -1)}.
\end{equation}
From these we derive for KIC~10685175  $i=76.53^{\circ}$ and $\beta=29.9^{\circ}$, or vice versa; the two angles are interchangeable as both pairs give the same pulsation amplitude and phase modulation with rotation. Pulsation maximum occurs when the pulsation pole is closest to the line of sight with an inclination of $i - \beta = 46.6^\circ$. Half a rotation cycle later, the secondary maximum occurs when the angle to the first pulsation pole is $i + \beta = 106.4^\circ$, or the angle to the other pole is $73.6^\circ$ to the line of sight. This secondary maximum occurs when the equatorial pulsation belt between pulsation latitudes $\pm 35.3^\circ$ dominates the visible hemisphere of the star. 

Returning to the pulsation phase as a function of rotation: The bottom panel of Figure\,\ref{fig:modulation} shows that the pulsation phase is poorly determined when the pulsation amplitude is low, since the uncertainty in the pulsation phase is inversely proportional to the pulsation amplitude \citep{1999DSSN...13...28M}. The pulsation phase does not show obvious periodic variations with rotation, but it does increase around rotation phase about 0.3 and 0.7. This happens when the pulsation amplitude goes to zero as the line-of-sight passes over a node, according to the oblique pulsator model. But the phase does not show the $\pi$-rad shift expected for a non-distorted pulsator, so this case is a distorted mode, but close to a pure quadrupole mode. 

\section{Spherical harmonic decomposition}

The pulsation amplitude and phase modulation of KIC~10685175 and the rotational light variations give information about the surface spots and pulsation axis geometry. For this distorted quadrupole mode, with $i=76.53^{\circ}$ and $\beta=29.9^{\circ}$ derived in the last section, the pulsation can be decomposed into the components of a spherical harmonic series ($\ell=0,\,1,\,2$) following the method of \citet{1992MNRAS.259..701K}. This was done using the frequencies, amplitudes, and phases from Table~2, and also the time zero point from Table~2, $t_0 = {\rm BJD}\,2458711.21931$, thus referring to the time of pulsation amplitude maximum. The results are shown in Table~3. 

\begin{deluxetable*}{ccccccc}
\tablenum{3}
\tablecaption{ Components of the spherical harmonic series description of the pulsation for $i=76.53^{\circ}$ and $\beta=29.9^{\circ}$ for the frequencies, amplitudes, and phases from Table~2, with $t_0 = {\rm BJD}\,2458711.21931$.}
\tablewidth{0pt}
\tablehead{
\colhead{ $\ell$ } & \colhead{$A^{(\ell)}_{-2}$ (mmag)} & \colhead{$A^{(\ell)}_{-1}$ (mmag)} & \colhead{$A^{(\ell)}_{0}$ (mmag)} & \colhead{$A^{(\ell)}_{+1}$ (mmag)} & \colhead{$A^{(\ell)}_{+2}$ (mmag)}  & \colhead{$\phi$ (rad)}
}
\startdata
2 & 0.189 & 0.315 & $-0.563$ & 0.315 & 0.189 & 0.518\\
1 &   & 0.032 & 0.027 & 0.033 & & $-1.952$\\
0 &   &   & 1.142 &   &   &0.636\\
\enddata
\end{deluxetable*}

From these results, the pulsation has very little dipole contribution ($\ell =1$) compared to quadrupole and radial contributions ($\ell =2,0$). The phases of the components $A_0^{(2)}$ and $A_0^{(0)}$ are nearly equal, indicating that, compared to a pure quadrupole mode, the amplitude is accentuated at the pulsation pole, and reduced at the equator. Let us look at the meaning of the components shown in Table~3. For the quadrupole components, the negative central component means that it is $\pi$\,rad out of phase with the other four components, which are in phase with each other, at the time of pulsation maximum, $t_0$. Thus, we can add these five amplitudes to find that at $t_0$  the quadrupole component is contributing 0.445\,mmag to the pulsation amplitude. Neglecting the insignificant dipole component then allows us to add the quadrupole amplitude to the radial component amplitude, since they are nearly in phase, to find an amplitude at pulsation maximum of 1.587\,mmag, in good agreement with Figure\,\ref{fig:modulation}. If we now do the same at rotation phase 0.5, the phases for the outer sidelobes are unchanged, since they beat twice per rotation, the phases for the inner side lobes are now negative, since they beat once per rotation, and the phase of the central frequency is unchanged. Thus the quadrupole contributes -0.815\,mmag; when added to the radial component, that then gives a pulsation amplitude at rotation phase of 0.5 of 0.327\,mmag, again in good agreement with Figure\,\ref{fig:modulation}. More precisely, a fit of all three spherical harmonic components taking into account that the exact phases seen in Table~3 gives the fit shown in Figure\,\ref{fig:modulation} as the red curves.

These results are consistent with the oblique pulsator model. The strong radial component of the decomposition suggests that the pulsation amplitude is  enhanced at the pulsation poles, and diminished at the pulsation equator, as can be expected with a dipolar magnetic field aligned with the pulsation mode. The pulsation phase changes are complicated. Those have been modelled by \citet{2018MNRAS.480.1676Q}, who show how those phase changes are a function of atmospheric depth and differ between the pole and equator as a consequence of the interaction of the acoustic and magnetic components of the pulsation in roAp star atmospheres.

\subsection{Other distorted quadrupole pulsators}

There are other roAp stars with pulsations that have been modelled with distorted quadrupoles using the technique of \citet{2005MNRAS.360.1022S}, such as  J1640 \citep{2018MNRAS.476..601H}, J1940 \citep{2018MNRAS.473...91H}, and HD~24355 \citep{2017EPJWC.16003004H}. We compare the simple derivation of $i$ and $\beta$ using equations 2 and 3 for these three stars with the values found in the listed papers using the models of Saio. The results are shown in Table~4. The agreement is very good for J1640 and J1940, and less so for HD~24355. We interpret this to mean that the modes are only mildly distorted quadrupole modes, and that the derivation of $i$ and $\beta$ from equations 2 and 3 is a good first approximation for the rotational inclination and magnetic obliquity in roAp stars pulsation in quadrupole modes. This is more powerful that the less restrictive constraint on those angles found for dipole modes, which only gives a value for $\tan i \tan \beta$, hence a range of $i$ and $\beta$.

\begin{deluxetable*}{ccc}
\tablenum{4}
\tablecaption{Comparison of i and $\beta$ derived through the method of \citet{1990MNRAS.247..558K} and the models of \citet{2005MNRAS.360.1022S}. All values are in degrees.}
\tablewidth{0pt}
\tablehead{
\colhead{ target}  & \colhead{(i,$\beta$) from this work} & \colhead{(i,$\beta$) from the previous work} 
}
\startdata
J1640 & (76.69, 12.96) & (70, 13)\\
J1940 &  (28.5, 83.65) & (30, 84)\\
HD~24355 & (26.24, 82.86) & (45, 77)\\
\enddata
\end{deluxetable*}

\section{Modeling quadrupole pulsation distorted by dipole magnetic fields}
\label{sec:modeling}

High-frequency axisymmetric pulsations of roAp stars are affected  by the presence of a magnetic field as discussed by, e.g., \citet{1996ApJ...458..338D}, \citet{2000MNRAS.319.1020C}, and \citet{2004MNRAS.350..485S}.
The surface amplitude distribution cannot be represented by a single Legendre function $P_\ell(\cos\theta_B)$, where $\theta_B$ is the angle measured from the magnetic axis; i.e., pure quadrupole (or dipole) pulsations cannot occur in a magnetic star.
In this section,  we obtain theoretical amplitude and phase modulations from the surface amplitude distributions calculated taking into account the magnetic effect, and compare them with those of  KIC\,10685175.
  
We obtain non-adiabatic pulsation eigenfunctions under  a dipole magnetic field using the method developed by Saio (2005), where pulsation variables are expressed by a sum of terms proportional to Legendre functions of different degrees; for example, the local luminosity  perturbation $\delta L$ is expressed as
\begin{equation}
\delta L(r,\theta_B,t) = e^{i\sigma t}\sum_{j=1}^K \delta L_j(r)N_{\ell_j}P_{\ell_j}(\cos\theta_B),
\end{equation}
where $\sigma$ is the angular frequency of pulsation, $N_{\ell_j}$ is a normarization factor, $\ell_j=2(j-1)$ (for even modes such as distorted quadrupole modes), and $K$ is the truncation length of the expansion ($K=10 - 20$ were adopted depending on the convergence of  eigenfunction).
The other five variables are also expressed in similar forms and substituted into linear nonradial pulsation equations including magnetic perturbations. The resulting $6K$ ($K$ is the truncation length of the expansion) dimensional differential equations with one complex eigenvalue $\sigma$  were solved for an equilibrium structure with an assumed value of $B_p$, polar strength of dipole magnetic field. 
The obtained latitudinal amplitude variation of $\delta L(R,\cos\theta_B)$ on the surface was converted to observational light variation at each rotation phase (see \citet{2004MNRAS.350..485S} for details) for the assumed obliquity and inclination angles $(\beta,i)$.
 
For the stellar structure model of KIC\,10685175 we adopted a model close to the ZAMS of $1.70\,{\rm M}_\odot$ with the initial chemical composition $(X,Z)=(0.70,0.02)$ in the fully ionized interior. The model has parameters; $\log L/{\rm L}_\odot= 0.896$, $\log T_{\rm eff}= 3.8918$ ($T_{\rm eff}= 7794$\,K), and $\log g= 4.29$ (cgs), which are consistent with those obtained by \citet{2019MNRAS.488...18H}; $T_{\rm eff}=8000 \pm 300$\,K, $\log L/{\rm L}_\odot= 0.896 \pm 0.022$, and $M/{\rm M}_\odot = 1.65 \pm 0.25$. In the outer layers, helium was assumed to be depleted to the He\,I ionization zone and convection was suppressed,  similarly to the polar model of \citet{2001MNRAS.323..362B}.  Similar envelope structures were adopted for other roAp stars such as J1640 \citep{2018MNRAS.476..601H}, J1940 \citep{2018MNRAS.473...91H} and HD24355 \citep{2017EPJWC.16003004H}.

\begin{figure}
\includegraphics[width=0.49\textwidth]{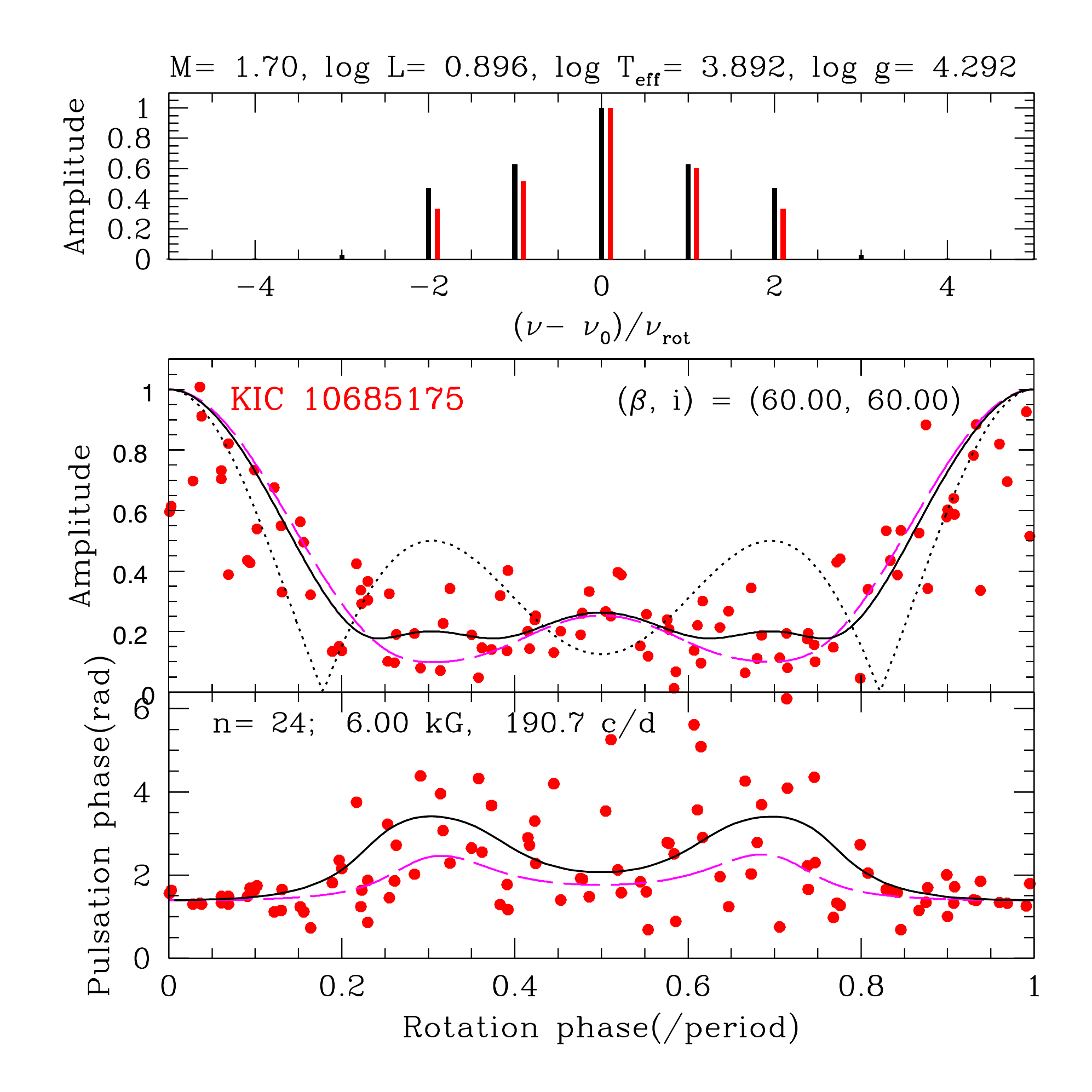}
\includegraphics[width=0.49\textwidth]{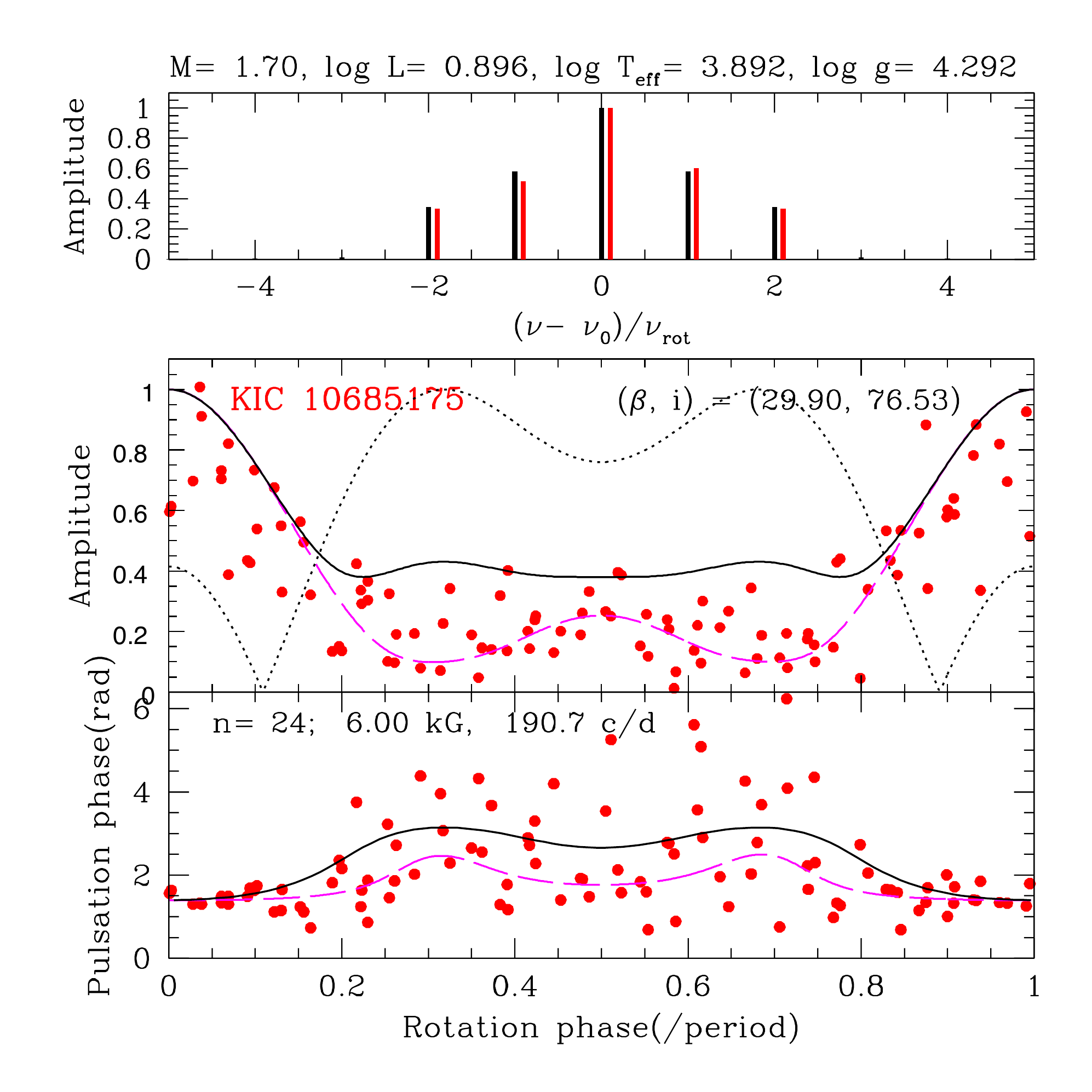}
\caption{Theoretical rotational sidelobes (top panels), amplitude (middle) and phase (bottom) modulations are compared with KIC\,10685175. Black and red colors are used for theoretical predictions and observation, respectively. Dotted lines in the middle panels show amplitude variations expected from a pure quadrupole mode. Dashed magenta lines correspond to red lines in Figure\,\ref{fig:modulation}.}
\label{fig:mod}
\end{figure}

We obtained eigenfunctions of an axisymmetric quadrupole p mode whose frequency is similar to the observed one, assuming various values of $B_{\rm p}$.  Then, we varied the obliquity $\beta$ and inclination $i$  to see whether  amplitude/phase modulations and the amplitudes of rotational sidelobes are consistent with the observed ones, while the limb darkening parameter $\mu$ was fixed at $0.6$ (results are insensitive to $\mu$). We examined eigenfunctions obtained assuming $B_{\rm p}$ ranging from 1 to 10\,kG, and found  that $B_{\rm p}$ must be as high as about 6\,kG to fit with observed amplitude/phase modulations and sidelobe amplitudes. The best case we obtained is shown in the left panel of Figure\,\ref{fig:mod}, for which $(B_{\rm p},\beta, i) = (6.0\,{\rm kG},60^\circ,60^\circ)$ were adopted. Because of the strong magnetic field, the eigenfunction is considerably distorted from a pure quadrupole mode. The kinetic energy of the $\ell=0$ component is about 10\% of the $\ell=2$ component, while the imaginary parts of both components are comparable with corresponding real parts indicating that the importance of the $\ell=0, 2$ components depends on the pulsation phase. If such a high magnetic field is present in KIC\,10685175, Zeeman splittings should appear in a high resolution spectrum of this star, hence can test this model.

The right panel of Figure\,\ref{fig:mod} shows amplitude/phase modulations and sidelobe amplitudes for the case with the same eigenfunction at $B_{\rm p} = 6$\,kG, but for $(\beta,i)= (29.9^\circ, 76.53^\circ)$, which were obtained in \S3.3.
The predictions are considerably different from the dashed lines, probably due to strong magnetic distortion of eigenfunctions.

\section{Discussion and Conclusions} 
\label{sec:conclusion}

Analysis of the TESS data confirms that KIC~10685175 is a roAp star, as was discovered with the super-Nyquist method from the {\it Kepler} long cadence data. The pulsation frequency of KIC~10685175 derived in this work, 191.5151\,d$^{-1}$, differs from that found by \citet{2019MNRAS.488...18H}, 240.45189\,d$^{-1}$, based on the {\it Kepler} long cadence data. Thus, in this previous work, the real frequency deduced is a Nyquist alias. Although the super-Nyquist peaks can be seen in the periodogram of the under-sampled {\it Kepler} long cadence data for this star, the frequency with the highest amplitude is not in this case the real pulsation frequency. A reanalysis of the Kepler data shows that this is simply the result of the signal-to-noise ratio: The alias found by \citet{2019MNRAS.488...18H} and the true frequency found by us in this work are closer  in amplitude in the {\it Kepler} data that the height of the noise peaks in the amplitude spectrum, so the noise in this case caused confusion for the {\it Kepler} data. We also derived the rotation frequency for KIC~10685175 to be  $0.32229 \pm 0.0005$\,d$^{-1}$ from the TESS data, which is consistent with the higher precision value of $P_{\rm rot} = 3.101988 \pm 0.000008$\,d found by \citet{2019MNRAS.488...18H} from the {\it Kepler} data.  

We inspected the pulsation amplitude and phase modulation caused by rotation. The double wave rotational light variation indicates that there are two spots on the surface. Two peaks with unequal height in pulsation amplitude modulation indicate that one pulsation pole and the equator dominate the pulsation amplitude over one rotation cycle. The pulsation modulation and pulsation phases of the sidelobes show that KIC~10685175 is a weakly distorted quadrupole pulsator. We calculated the rotation inclination, $i$, and magnetic obliquity, $\beta$, following \citet{1990MNRAS.247..558K}, which provided detailed information of geometry and we used those with a spherical harmonic decomposition to better understand the pulsation geometry and the distortion from a pure quadrupole mode. These angles show that both pulsation (magnetic) poles are seen over the rotation cycle, which is consistent with the double wave light variations caused by the polar spots. 

To understand the pulsation, we have modelled the theoretical amplitude and phase modulation of this star and found that the polar magnetic field strength is as high as about 6\,kG which is predicted to be observed in the high resolution spectrum of this star. If a similar magnetic field can be detected in this star, it would be a critical test to this model. However, there are considerable differences between the observation and the theoretic model with the same eigenfunction at $B_{\rm p} = 6$\,kG, but for $(\beta,i)= (29.9^\circ, 76.53^\circ)$, which are obtained in \S3.3. This is probably due to strong magnetic distortion of eigenfunctions. The best fit model gives us the $(B_{\rm p},\beta, i) = (6.0\,{\rm kG},60^\circ,60^\circ)$, and also stellar parameters that are consistent with those obtained by \citet{2019MNRAS.488...18H}.

\section*{Acknowledgements}
This work was funded by the National Natural Science Foundation of China (NSFC) under grants No.11973001, No.11833002, No.11673003, and National KeyR$\&$D Program of China under grant No.2019YFA0405500.
This work includes data collected by the TESS mission. Funding for the TESS mission is provided by the NASA Explorer Program.
This work has made use of data products from the Guoshoujing Telescope (the Large Sky AreaMulti-Object Fibre Spectroscopic Telescope, LAMOST). LAMOST is a National Major Scientific Project built by the Chinese Academy of Sciences. Funding for the project has been provided by the National Development and Reform Commission. LAMOST is operated and managed by the National Astronomical Observatories, Chinese Academy of Sciences.

\bibliography{16ver_10685175}{}
\bibliographystyle{aasjournal}

\end{document}